\documentclass{article}
\usepackage{eurosym}
\usepackage[T1]{fontenc}
\usepackage{epsf,epsfig,subfigure}
\usepackage{float,color,ulem}
\usepackage{amssymb}
\usepackage{amsfonts,mathrsfs}
\usepackage{amsmath}
\usepackage[english]{babel}
\usepackage{graphicx,multirow}
\usepackage{indentfirst}
\usepackage{bbm}
\usepackage{geometry}
\usepackage[notref,notcite]{showkeys}
\graphicspath{{./Pictures/}}
\usepackage[latin1]{inputenc}
\usepackage{listings}

\usepackage{paralist,graphics,epsfig,graphicx,epstopdf,mathrsfs}
\usepackage{float,color,ulem,comment,tabulary,booktabs}

\newcommand\doubleRule{\toprule\toprule}

\newfloat{figure}{H}{lof}
\newfloat{table}{H}{lot}
\floatname{figure}{\figurename}
\floatname{table}{\tablename}
\usepackage[notref,notcite]{showkeys}

\newfloat{figure}{H}{lof}
\newfloat{table}{H}{lot}
\floatname{figure}{\figurename}
\floatname{table}{\tablename}

\numberwithin{equation}{section}

\renewcommand{\eqref}[1]{(\ref{#1})}

\begin{document}

\title{\textbf{Predicting the cumulative number of cases for the COVID-19 epidemic in  China from early  data}}
\author{\textsc{Z. Liu$^{(a)}${\small \thanks{Research was partially supported by NSFC and CNRS (Grant Nos. 11871007 and 11811530272) and the Fundamental Research Funds for the Central Universities.}} \thanks{Corresponding authors.},
P. Magal$^{(a,b)}${\small \thanks{Research was partially supported by CNRS and  National Natural Science Foundation of China (Grant No.11811530272)}} $^{ \dag }$ , O. Seydi$^{(c)}$, \text{ and} G. Webb$^{(d) \dag}$   }\\
$^{(a)}${\small \textit{School of Mathematical Sciences, Beijing Normal University,}}\\
{\small \textit{Beijing 100875, People's Republic of China }} \\
$^{(b)}${\small \textit{Univ. Bordeaux, IMB, UMR 5251, F-33400 Talence, France.}} \\
{\small \textit{CNRS, IMB, UMR 5251, F-33400 Talence, France.}}\\
$^{(c)}${\small \textit{D$\acute{e}$partement Tronc Commun, $\acute{E}$cole Polytechnique de Thi$\grave{e}$s, S$\acute{e}$n$\acute{e}$gal}} \\
$^{(d)}${\small \textit{Mathematics Department, Vanderbilt University, Nashville, TN, USA}}
}
\maketitle

\begin{abstract}
We model the COVID-19 coronavirus epidemic in China. We use early reported case data to predict the cumulative number of reported cases to a final size. The key features of our model are the timing of implementation of major public policies restricting  social movement, the identification and isolation of unreported cases, and the impact of asymptomatic infectious cases.
\end{abstract}

\section{Introduction}

Many mathematical models of the COVID-19 coronavirus epidemic in China have been developed, and some of these are listed in our references \cite{Hui, NLA1,ROO,SW, TBLTXW, TWLBTXW, T, WLL, Z}.  We develop here a model describing this  epidemic, focused on the effects of the Chinese government imposed public policies designed to contain this epidemic, and the number of reported and unreported cases that have occurred.  Our model here is based on our model of this epidemic in \cite{LMSW}, which was focused on  the early phase of this epidemic (January 20 through January 29) in the city of Wuhan, the epicenter of the early outbreak. During this early phase, the cumulative number of daily reported cases grew exponentially. In \cite{LMSW}, we identified a constant transmission rate corresponding to this exponential growth rate of the cumulative reported cases, during this early phase in Wuhan.  

On January 23, 2020,  the Chinese government imposed major public restrictions on the population of Wuhan. Soon after, the epidemic in Wuhan  passed beyond the early exponential growth phase, to a phase with slowing growth. In this work, we assume that these major government measures caused the transmission rate to change from a constant rate to a time dependent exponentially decreasing rate. We identify this exponentially decreasing transmission  rate based on reported case data after January 29. We then  extend our model of the epidemic to the central region of China, where most cases occurred. Within just a few days after January 29, our model can be used to project the time-line of the model  forward in time, with increasing accuracy, to a final size.

\section{Model}

The model consists of the following system of ordinary differential equations:
\begin{equation}\label{2.1}
\begin{cases}
S'(t)=-\tau(t) S(t)[I(t)+U(t)],\\
I'(t)=\tau(t) S(t)[I(t)+U(t)]- \nu I(t),\\
R'(t)=\nu_1 I(t)-\eta R(t),\\
U'(t)=\nu_2 I(t)-\eta U(t).\\
\end{cases}
\end{equation}
This system is supplemented by initial data
\begin{equation}\label{2.2}
S(t_0)=S_0>0,\, I(t_0)=I_0>0,\, R(t_0)=0\text{ and }U(t_0)=U_0 \geq 0.
\end{equation}
Here $t \geq t_0$ is time in days, $t_0$ is the beginning date of the model of the epidemic,
$S(t)$ is the number of individuals susceptible to infection at time $t$,
$I(t)$ is the number of asymptomatic infectious individuals at time $t$,
$R(t)$ is the number of reported symptomatic infectious individuals at time $t$, and
$U(t)$ is the number of unreported symptomatic infectious individuals at time $t$.

Asymptomatic infectious individuals $I(t)$ are infectious for an average period of $1 / \nu$ days. Reported  symptomatic individuals $R(t)$ are infectious for an average period of $1 / \eta$ days, as are unreported symptomatic individuals
$U(t)$. We assume that reported symptomatic infectious individuals $R(t)$ are reported and isolated immediately, and cause no further infections.
The asymptomatic individuals $I(t)$ can also  be viewed as having a low-level symptomatic state. All infections are acquired from either $I(t)$ or $U(t)$ individuals. 

The parameters of the model are listed in Table \ref{Table1} and a schematic diagram of the model is given in Figure 1.

\vspace{0.2in}	

\begin{table}[H] \centering
		\begin{footnotesize}
		\begin{tabular}{ccccc}
			\doubleRule
			\textbf{Symbol}  &  \textbf{Interpretation} & \textbf{Method}     \\
			\hline
			\hline
			$ t_0 $
			& {\footnotesize Time at which the epidemic started}
			& fitted
			\\
			$S_0$
		    & {\footnotesize Number of susceptible at time $t_0$ }
		    & fixed
			\\
			$I_0$
		    & {\footnotesize Number of asymptomatic infectious at time $t_0$ }
		    & fitted
		    \\
			$U_0$
		    & {\footnotesize Number of unreported symptomatic infectious at time $t_0$ }
		    & fitted
		    \\				
			$ \tau(t) $
			&Transmission rate at time $t$
			&
			fitted
			&
			\\
			$1/\nu $
			& Average time during which asymptomatic infectious are asymptomatic
			& fixed
			\\
			$f$
			& Fraction of asymptomatic infectious that become reported symptomatic infectious
			& fixed
			\\
			$ \nu_1=f\, \nu$
			&  Rate at which asymptomatic  infectious become reported symptomatic
			& fitted
			\\
			$ \nu_2=(1-f)\, \nu$
			&  Rate at which asymptomatic infectious become unreported symptomatic
			& fitted
			\\
			$1/ \eta$
			& Average time symptomatic infectious have symptoms
			& fixed
			\\
			\doubleRule
		\end{tabular}
	         \end{footnotesize}
		\caption{\textit{Parameters of the model.}}\label{Table1}
	\end{table}
	
\vspace{0.2in}	

\begin{figure}
\begin{center}
\includegraphics[scale=0.7]{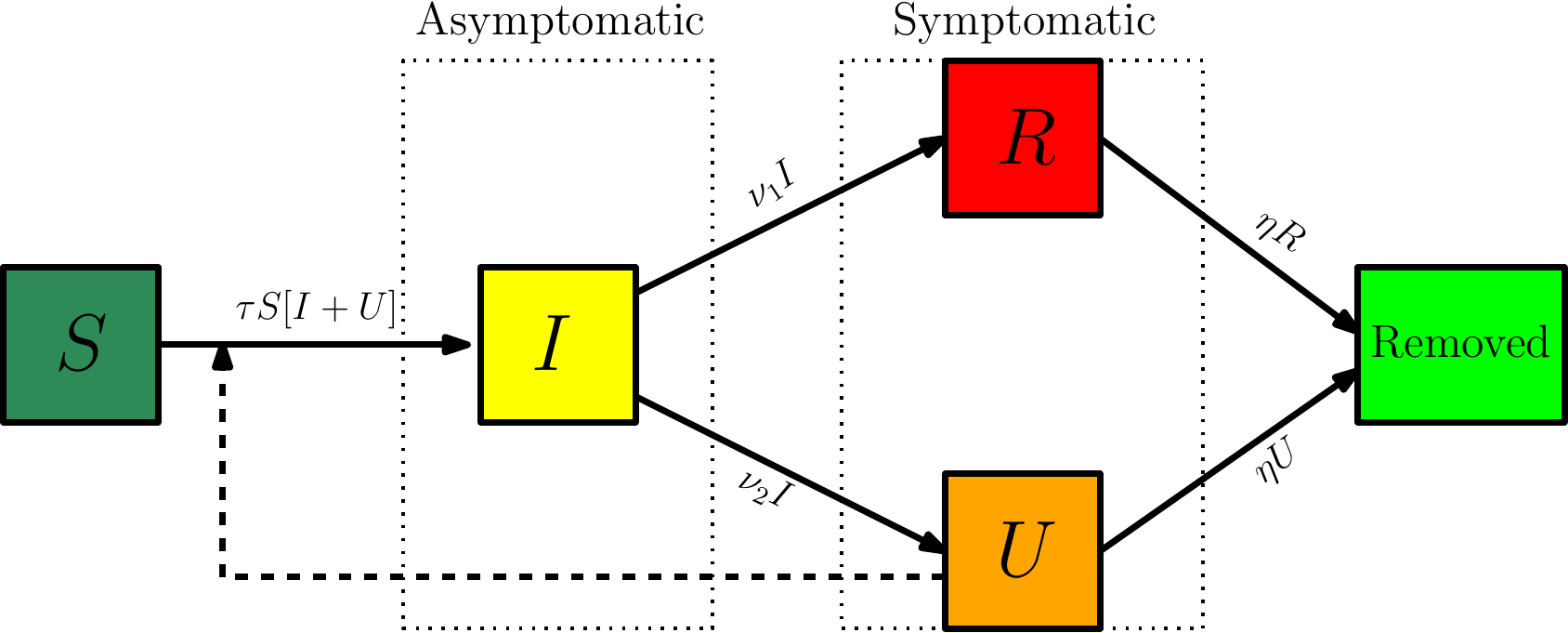}
\end{center}
\caption{\textit{Compartments and flow chart of the model.}}
\label{fig1}
\end{figure}
	
\section{Data}

We use data from the National Health Commission of the People's Republic of China and the Chinese CDC for mainland China as of February 15, 2020:

\begin{table}[H]
\centering
\begin{footnotesize}
\begin{tabular}{lcccccccccccc}
&    \\ \hline
January			& 20 & 21 &  22 & 23 & 24 & 25 & 26 & 27 & 28 & 29 &3 0 & 31 \\ \hline
 				& $291$ & $440$ & $571$  & $830$ & $1287$ & $1975$ & $2744$ & $4515$ & $5974$ & $7711$  & $9692$ &  $11791$ \\
                                 & & & & & & & &  &  & \\

February                   & 1 & 2 &  3 & 4 & 5 & 6 & 7 & 8 & 9 & 10 & 11 & 12  \\ \hline
                                 & $14380$ & $17205$ & $20438$  & $24324$ & $28018$ & $31161$ & $34546$ &  $37198$ & $40171$ & $42638$ & $44653$ &  $46472$ \\
                                 & & & & & & & &  &  & \\

February                   & 13 & 14 & 15   \\ \hline

                                  & $48467$ &  $49970$ & $51091$ \\
\hline
\end{tabular}
\end{footnotesize}
\caption{\textit{Cumulative daily reported case data from January 20, 2020 to February 15, 2020, reported for mainland China
by the National Health Commission of the People's Republic of China and the Chinese CDC.}}
\label{Table2}
\end{table}

We plot the data for daily reported cases and the cumulative reported cases in Figure 2.

\begin{figure}
\begin{center}
\includegraphics[scale=0.65]{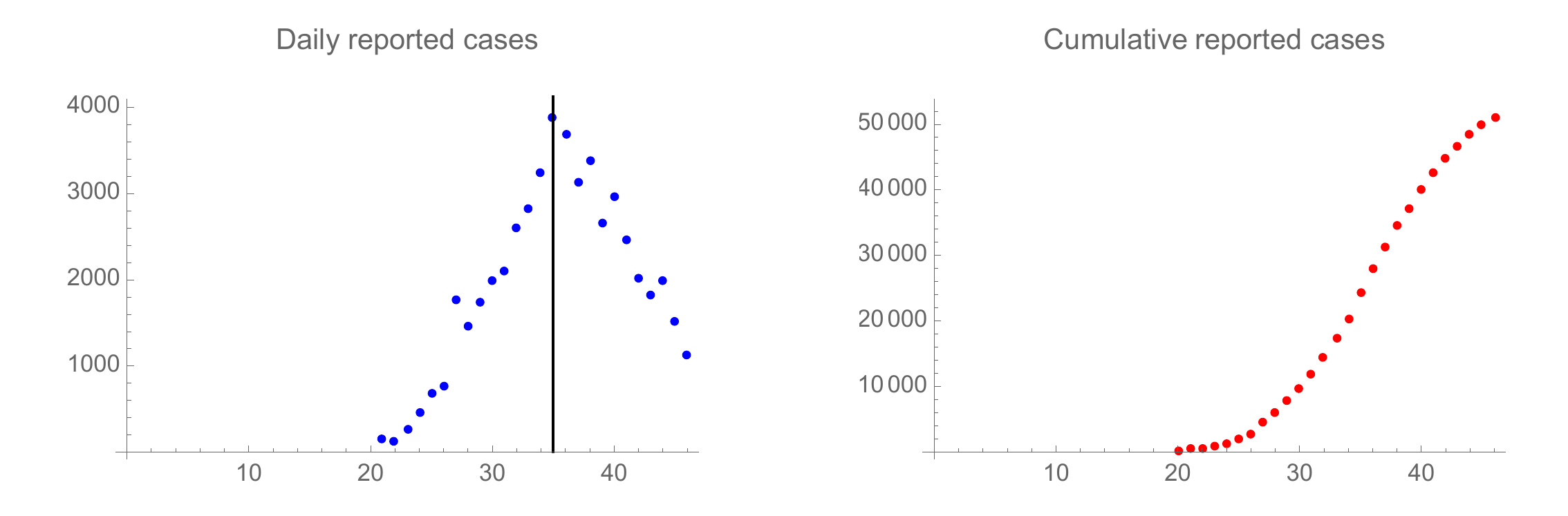}
\end{center}
\caption{\textit{Daily reported cases data (left) and cumulative reported cases data (right). The epidemic turning point of the reported case data is approximately February 4, 2020 (day 35, day 1 = January 1, 2020).}}
\label{fig2}
\end{figure}

\section{Model parameters}

We assume $f = 0.8$, which means that $20\%$ of symptomatic infectious cases go unreported. We assume $\eta =1/7$, which means that the average period of infectiousness of both unreported symptomatic infectious individuals and reported symptomatic infectious individuals is $7$ days. We assume $\nu =1/7$, which means that the average period of infectiousness of asymptomatic infectious individuals is $7$ days. These values can be modified as further epidemiological information becomes known.

In our previous work, we assumed that in the early phase of the epidemic (January 20 through January 29), the cumulative number  of reported cases grew approximately exponentially, according to the formula:
$$CR(t) = \chi_1 \exp(\chi_2 t) - \chi_3, \, \, t \geq t_0$$
with values $\chi_1=0.16$,
$\chi_2=0.38$,
$\chi_3=1.1$.
These values of $\chi_1$, $\chi_2$, and $\chi_3$ were fitted to reported case data from January 20 to January 29.
We assumed the initial value $S_0 = 11,000,000$, the population of the city Wuhan, which was the epicenter of the epidemic outbreak.
The other  initial conditions  are
$$I_0 = \frac{\chi_2 \chi_3}{f (\nu_1 + \nu_2)} = 3.7, \, \, \,
U_0 = \bigg(\frac{(1 - f) (\nu_1 + \nu_2)}{\eta + \chi_2}\bigg) \, I_0 = 0.2,\, \, \,
R_0 = 0.0.$$
The value of the transmission rate $\tau(t)$, during the early phase of the epidemic, when the cumulative number  of reported cases was approximately exponential, is the constant value
$$\tau_0=\bigg(\frac{\chi_2 \, + \nu_1 + \nu_2}{S_0}\bigg) \, \bigg(\frac{\eta \, + \, \chi_2}{\nu_2 \, + \, \eta  +  \, \chi_2}\bigg) \, = \, 4.47 \times 10^{-8}.$$
The initial time is
$$ t_0 \, =5  \, = \frac{1}{\chi_2} \, \bigg( \log(\chi_3) \, - \, \log(\chi_1) \bigg).$$
The value of the basic reproductive number is
$${\cal R}_0 \, = \,\bigg(\frac{\tau_0 S_0}{\nu_1 + \nu_2}\bigg) \, \bigg(1 \, + \, \frac{\nu_2}{\eta}\bigg) \,= \, 4.16.$$
These parameter formulas were derived in \cite{LMSW}.

After January 23,  strong government measures in all of China, such as isolation, quarantine, and public closings,   strongly  impacted the transmission of new cases. The actual effects of these measures were complex, and we use an exponential decrease for the transmission rate $\tau(t)$ to incorporate these effects after the early exponential increase phase.
The formula for $\tau(t)$ during the  exponential decreasing phase was derived by a fitting procedure. The formula for
$\tau(t)$ is
\begin{equation}
\begin{cases}
\tau(t) = \tau_0,  \, 0 \leq  \, t \, \leq \, 24,\\
\tau(t) = \tau_0 \, \exp \left(- \mu \left( t-24 \right) \right), \, 24 <t,
\end{cases}
\end{equation}
where January 24 and $\mu = 0.12$ are fitted from on-going reported case data after January 24. In Figure 3, we plot the graph of $\tau(t)$.

\begin{figure}
\begin{center}
\includegraphics[scale=0.7]{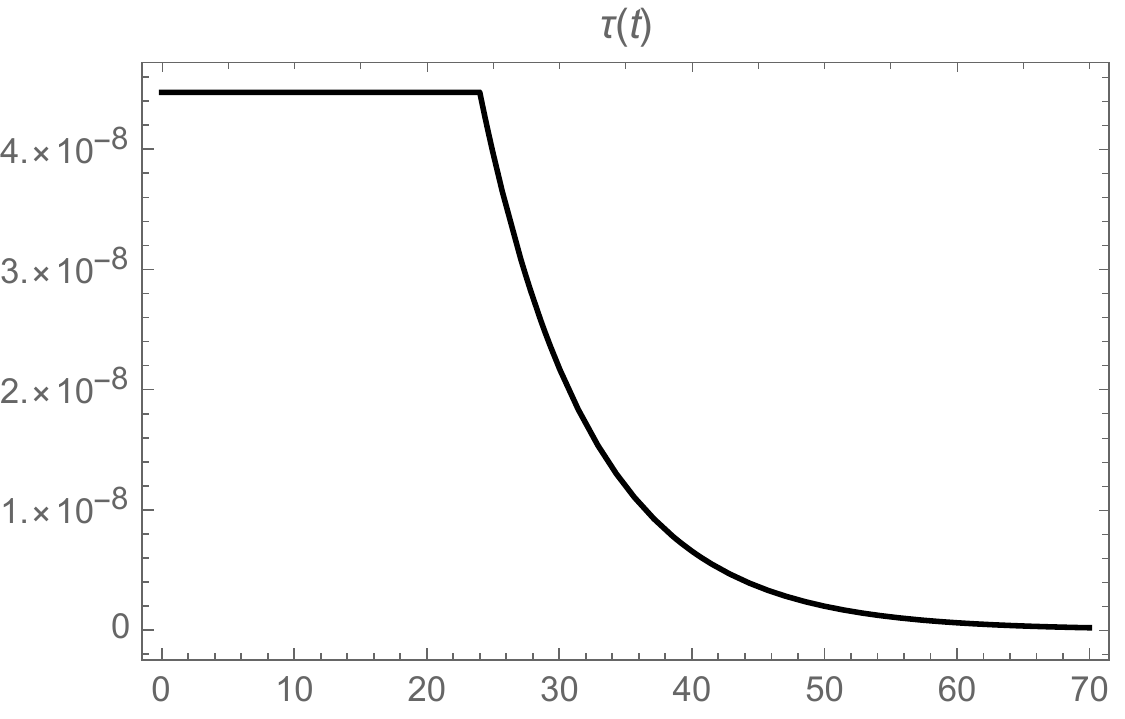}
\end{center}
\caption{\textit{Graph of the transmission rate  $\tau(t)$.}}
\label{fig3}
\end{figure}

\section{Model simulation}

We numerically simulated the model (2.1) to project forward in time the time-line of the epidemic after the government imposed interventions. 
We set $\tau(t)$ in (4.1) to $S_0  \, \tau(t) / 1,400,050,000$, where $S_0 = 11,000,000$ and  $1,400,050,000$ is the population of  mainland China, excluding Hong Kong, Macao and Taiwan.
We set $S(t_0)$ in (2.2) to $1,400,050,000$.
We set $t_0 = 5.0$, $I(t_0) = 3.7$, $U(t_0) = 0.2$ and $R(t_0) = 0$.
In Figure 4, we plot the graphs of $CR(t)$ (cumulative reported cases), $CU(t)$ (cumulative unreported cases), $R(t)$,  and $U(t)$ from the numerical simulation.
\begin{figure}
\begin{center}
\includegraphics[scale=0.7]{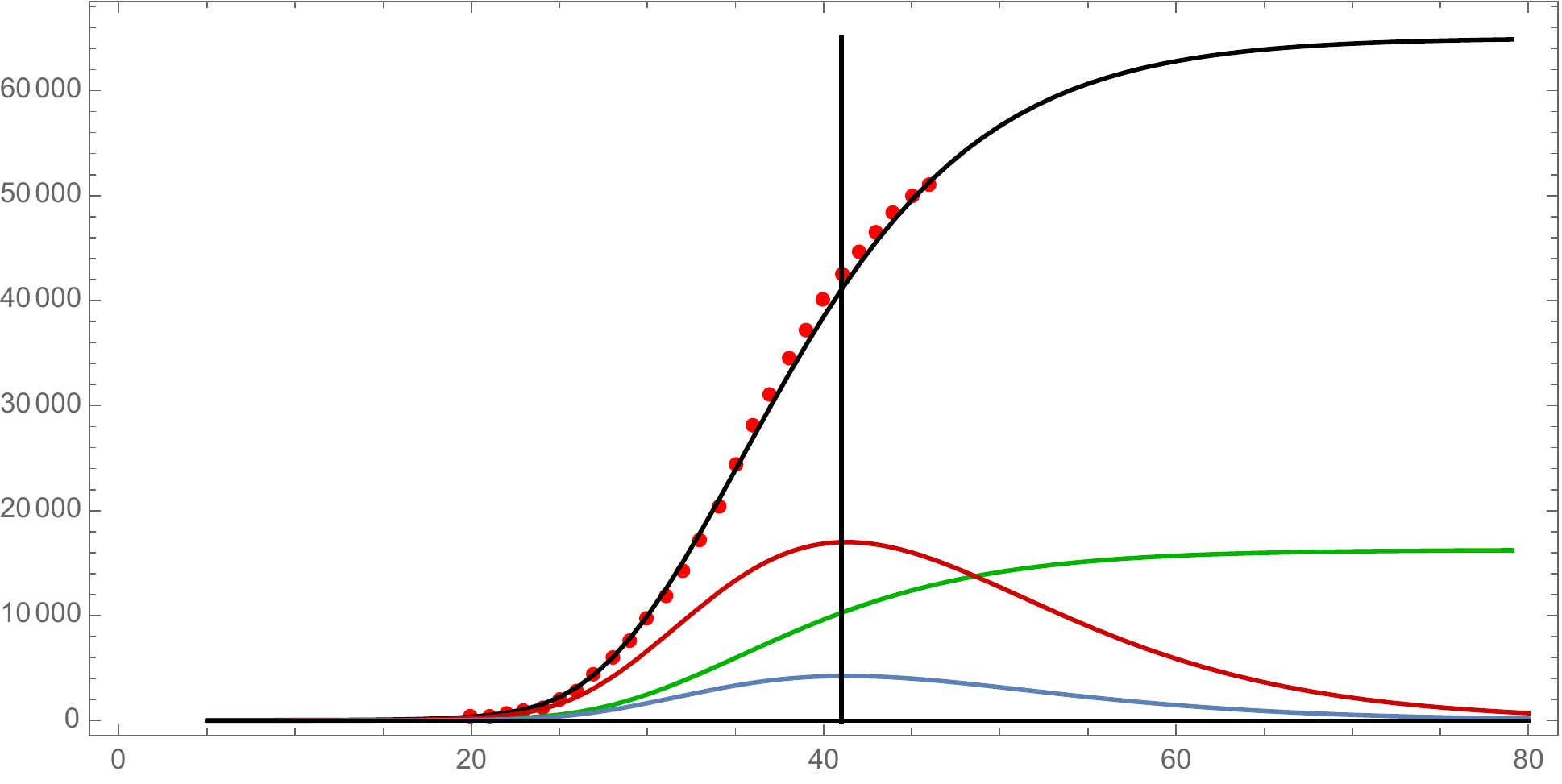}
\end{center}
\caption{\textit{Graphs of $CR(t)$ (black), $CU(t)$ (green), $U(t)$ (blue) and $R(t)$ (red). The red dots are the reported case data. The final size of the epidemic is approximately $65,000$  reported cases, approximately $16,000$ unreported cases, and approximately $81,000$ total cases.  The basic reproductive number is ${\cal R}_0 = 4.16$. The turning point of the epidemic is approximately  day 41 = February 10.}}
\label{fig4}
\end{figure}

In Figure 5 we plot the graphs of the reported cases $R(t)$ and  the infectious pre-symptomatic cases $I(t)$. The blue dots are obtained from the reported cases data (Table 2) for each day beginning on  January 26, by subtracting from each day, the value of the reported cases one week earlier.

\begin{figure}
\begin{center}
\includegraphics[scale=0.6]{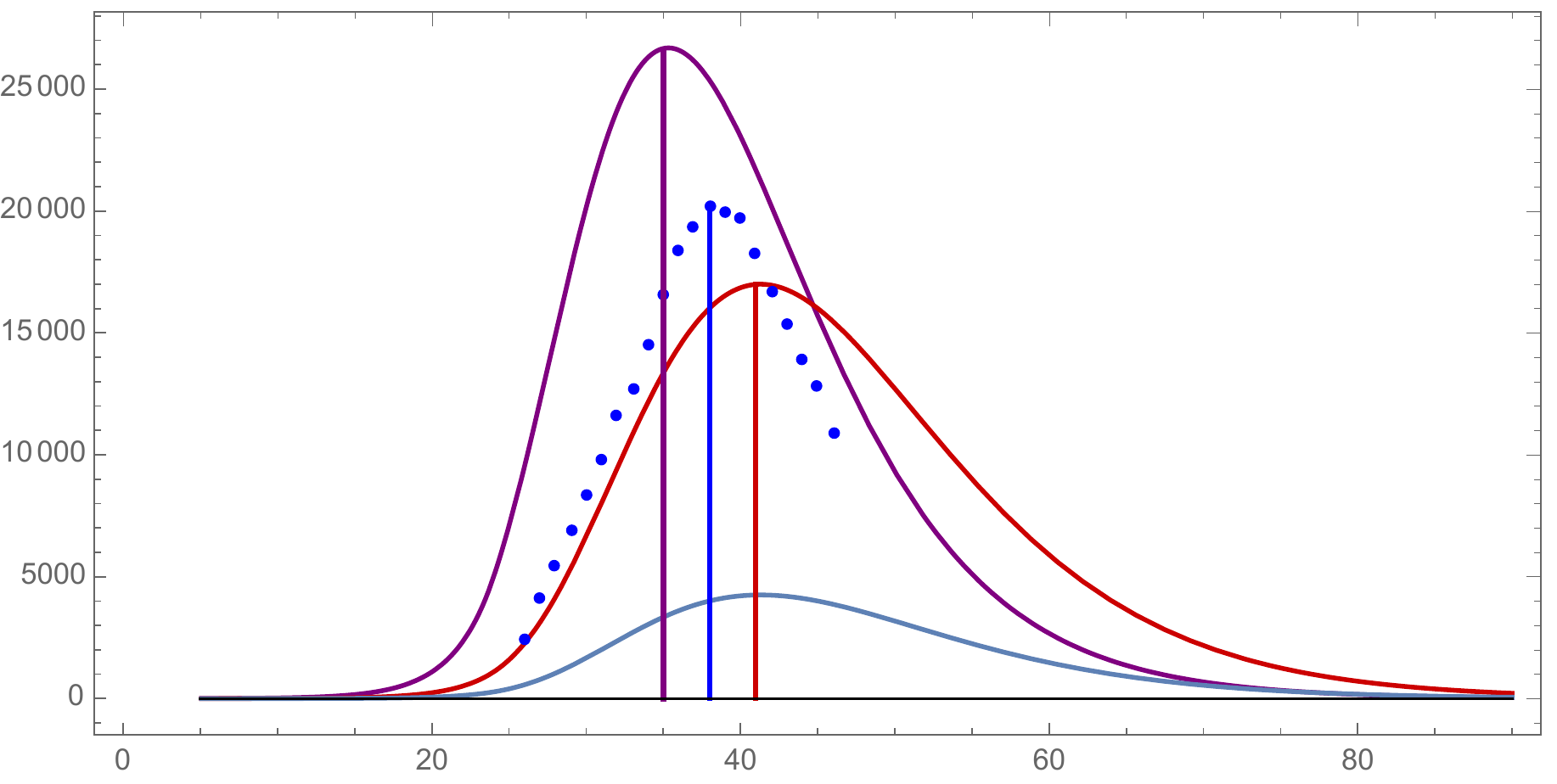}
\end{center}
\caption{\textit{Graphs of $R(t)$ (red), $U(t)$ (blue), and $I(t)$ (purple). The blue dots are the day by day weekly reported data.
The turning point of the asymptomatic infectious cases $I(t)$ is approximately day 35. The turning point of the reported cases $R(t)$ 
and the unreported cases $U(t)$ is approximately day 41. The turning point of the day by day weekly reported data is approximately day 38.}}
\label{fig5}
\end{figure}

Our model transmission rate $\tau(t)$ can be modified to illustrate the effects of an earlier  or later implementation of the major public policy interventions that occurred in this epidemic. The implementation one week earlier (24 is replaced by 17 in (4.1)) is graphed in Figure 6 (top).
All other parameters and the initial conditions remain the same. The total reported cases is approximately $4,500$  and the total unreported cases is approximately $1,100$. The implementation one week later (24 is replaced by 31 in (4.1)) is graphed in Figure 6 (bottom). The total reported cases is approximately $820,000$  and the total unreported cases is approximately $200,000$. The timing of the institution of major social  restrictions is critically important in mitigating the epidemic.

\begin{figure}
\begin{center}
\includegraphics[scale=.9]{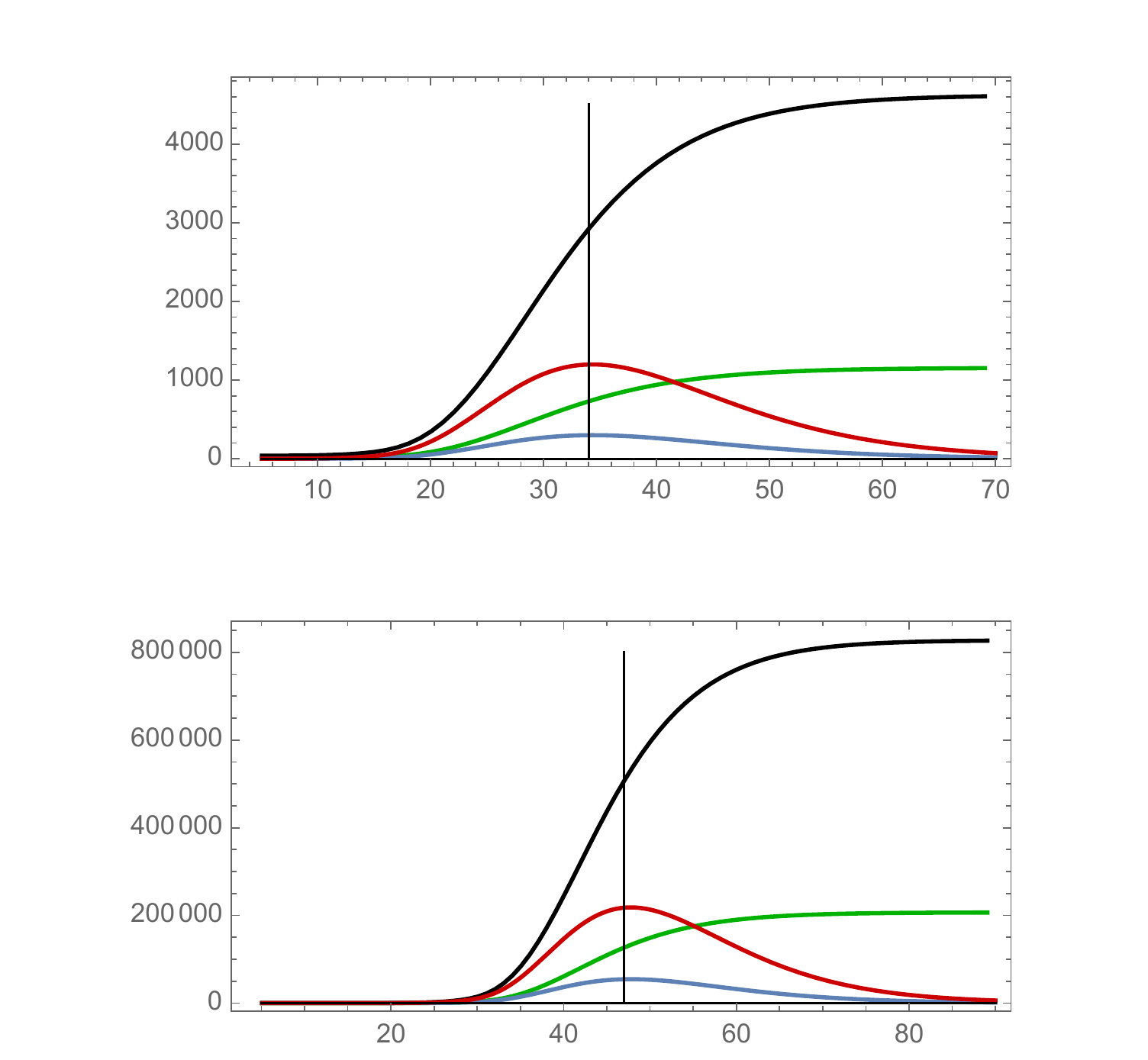}
\end{center}
\caption{\textit{Graphs of $CR(t)$ (black), $CU(t)$ (green), $U(t)$ (blue), and $R(t)$ (red). Top: The major public policy interventions were implemented one week earlier (January 17). Bottom: The major public policy interventions were implemented one week later (February 1).
The one week earlier implementation resulted in a final size of approximately $5,750$ total cases, with turning point day 34 = February 3.
The one week later implementation resulted in a final size of approximately $1,234,000$ total cases, with turning point day 47 = February 16.}}
\label{fig6}
\end{figure}

The number of unreported cases is of major importance in understanding the evolution of an epidemic, and involves great difficulty in their estimation.  The data from January 20 to February 15 for  reported cases in Table 2, was only for  tested cases. Between February 11 and February 15, additional clinically diagnosed case data, based on medical imaging showing signs of pneumonia, was also reported by the Chinese CDC. Since February 16, only tested case data has been reported by the Chinese CDC, because new NHC guidelines removed the clinically diagnosed category. Thus, after February 15, there is a gap in the reported case data that we used up to February 15. The uncertainty of the number of unreported cases for this epidemic includes this gap, but goes even further to include additional unreported cases. 

We assumed previously that the fraction $f$ of reported cases was $f = 0.8$ and the fraction of unreported cases was $1-f = 0.2$.  
Our model formulation can be applied with varying values for the fraction $f$. 
In Figure 7, we provide illustrations with the fraction $f = 0.4$ (top) and  $f = 0.6$ (bottom). 
The formula for the time dependent transmission rate $\tau(t)$ in (4.1) involves a new value for $\tau_{0}$ and $\mu$ for each case. The final size of the epidemic when $f = 0.4$ is approximately 164,700 cases, and the final size of the epidemic when $f = 0.6$ is approximately 110,700 cases.   From these simulations, we see that estimation of  the number of unreported cases has major importance in understanding the severity of this epidemic.

\begin{figure}
\begin{center}
\includegraphics[scale=.7]{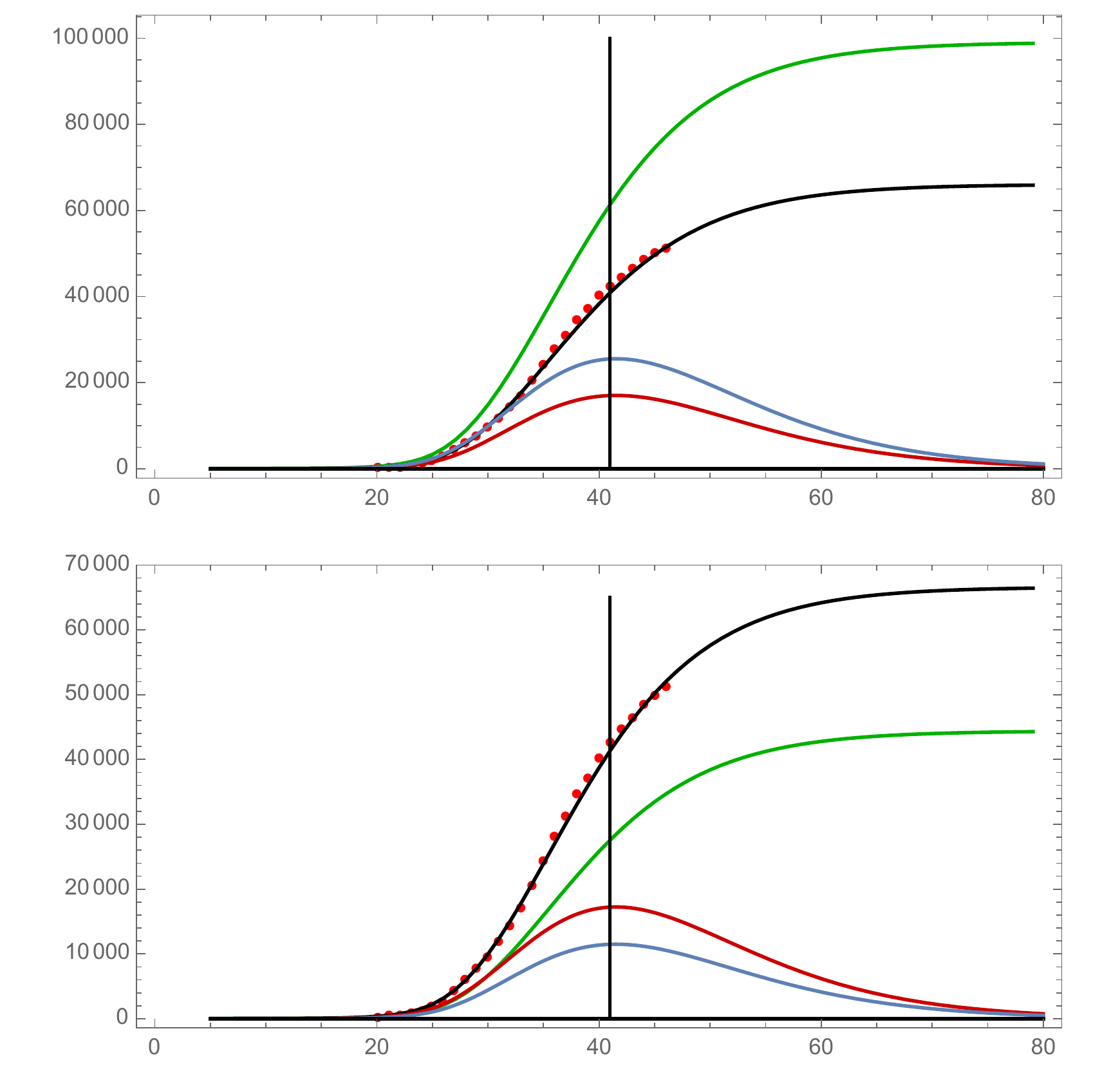}
\end{center}
\caption{\textit{Graphs of $CR(t)$ (black), $CU(t)$ (green), $U(t)$ (blue), and $R(t)$ (red).  The red dots are the cumulative reported 
case data from Table 2.
Top: $f = 0.4$, $t_0 = 5.0$, $\tau_0 = 4.054 \times 10^{-8}$, $\mu = 0.127$, $I(t_0)= 7.32$, $U(t_0) = 1.20$, $R(t_0) = 0.0$, and the basic reproductive number is ${\cal R}_0 = 5.03$.  The final size of the epidemic is approximately $65,900$  reported cases, approximately $98,800$ unreported cases, and approximately $164,700$ total cases. The turning point of the epidemic is approximately  day 41 = February 10. 
Bottom: $f = 0.6$, $t_0 = 5.0$, $\tau_0 = 4.254 \times 10^{-8}$, $\mu = 0.123$, $I(t_0) = 4.88$, $U(t_0) = 0.53$, $R(t_0) = 0.0$, and the basic reproductive number is ${\cal R}_0 = 4.25$.  The final size of the epidemic is approximately $66,500$ reported cases, approximately $44,300$ unreported cases, and approximately $110,800$ total cases. The turning point of the epidemic is approximately  day 41 = February 10.}}
\label{fig7}
\end{figure}

The number of days an asymptomatic infected individual is infectious is uncertain. We simulate in Figure 8 the model  with $\nu = 1/3$, which means that asymptomatic infected individuals are infectious on average 3 days before becoming symptomatic. The result is a small decrease in the final size of the epidemic, as compared to the case that $\nu = 1/7$.

\begin{figure}
\begin{center}
\includegraphics[scale=.6]{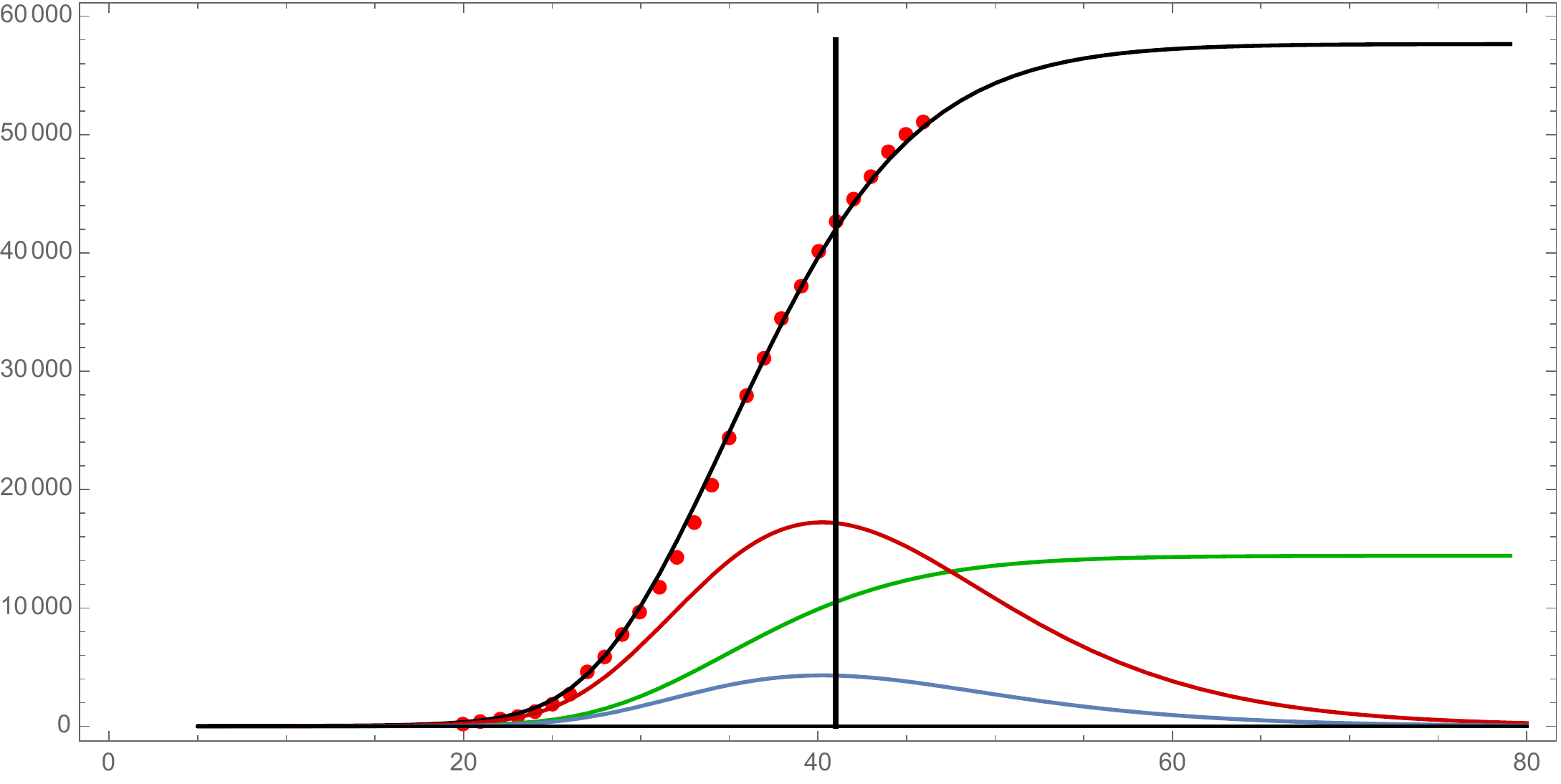}
\end{center}
\caption{\textit{Graphs of $CR(t)$ (black), $CU(t)$ (green), $U(t)$ (blue), and $R(t)$ (red).  The red dots are the cumulative reported 
case data from Table 2.
$f = 0.8$, $t_0 = 5.0$, $\tau_0 = 5.709 \times 10^{-8}$, $\mu = 0.082$, $I(t_0)= 1.57$, $U(t_0) = 0.2$, $R(t_0) = 0.0$, and the basic reproductive number is ${\cal R}_0 = 2.78$.  The final size of the epidemic is approximately $57,600$  reported cases, approximately $14,400$ unreported cases, and approximately $72,000$ total cases. The turning point of the epidemic is approximately  day 41 = February 10.}}
\label{fig8}
\end{figure}

We illustrate the importance of the level of government imposed public restrictions by decreasing the value of $\mu$ in formula (4.1) for 
$\tau(t)$. In Figure 9 we set $\mu = 0.09$ instead of $\mu = 0.12$, and the result is a significant increase in the number of cases.

\begin{figure}
\begin{center}
\includegraphics[scale=.6]{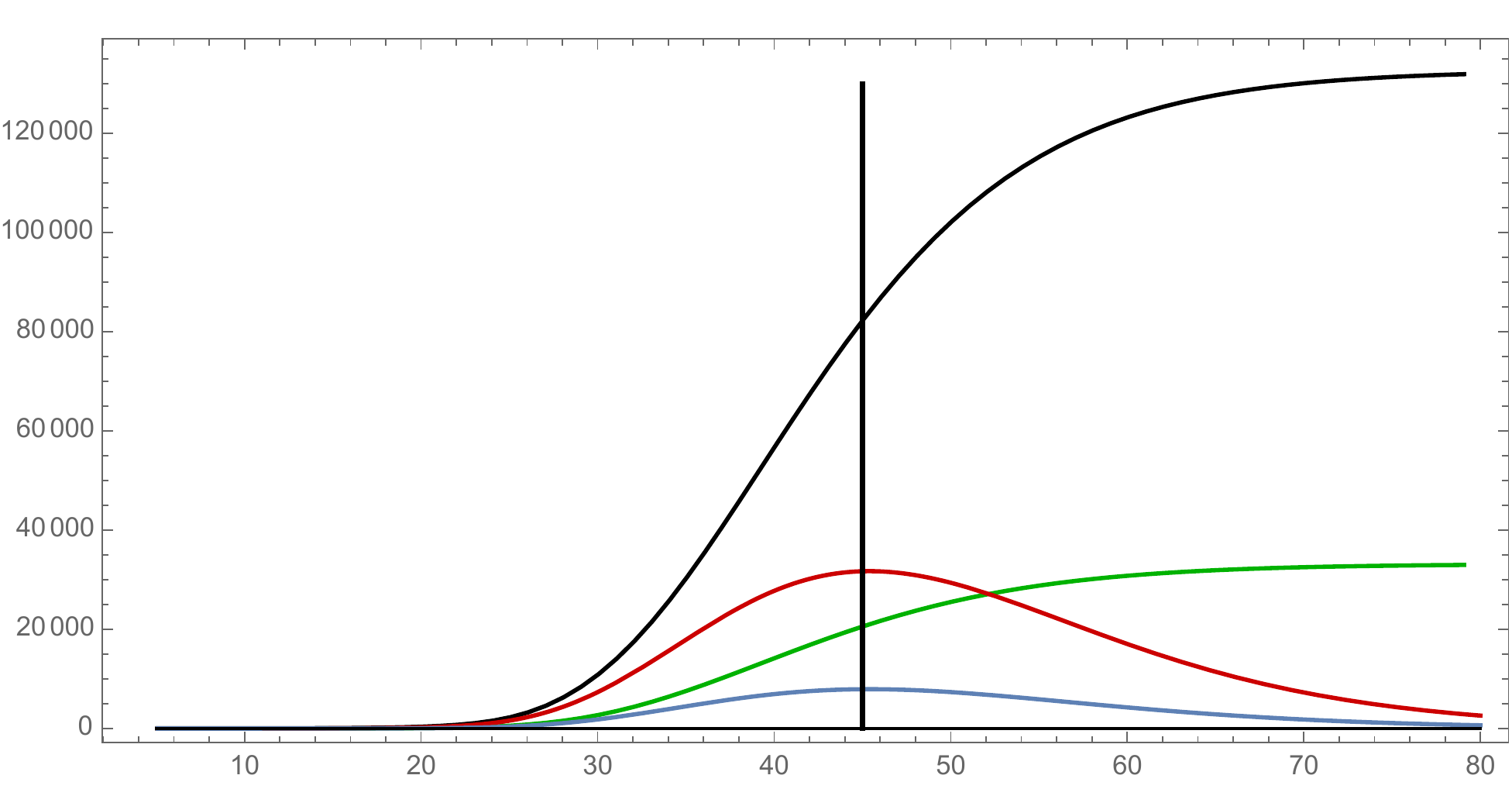}
\end{center}
\caption{\textit{Graphs of $CR(t)$ (black), $CU(t)$ (green), $U(t)$ (blue), and $R(t)$ (red).  The red dots are the cumulative reported 
case data from Table 2.
$f = 0.8$, $t_0 = 5.0$, $\tau_0 = 5.507 \times 10^{-8}$, $\mu = 0.09$, $I(t_0)= 3.66$, $U(t_0) = 0.2$, $R(t_0) = 0.0$, and the basic reproductive number is ${\cal R}_0 = 4.16$.  The final size of the epidemic is approximately $131,900$  reported cases, approximately $33,000$ unreported cases, and approximately $164,900$ total cases. The turning point of the epidemic is approximately  day 45 = February 14.}}
\label{fig9}
\end{figure}

\section{Discussion}

We have developed a model of the COVID-19 epidemic in China that incorporates key features of this epidemic:
(1) the importance of the timing and magnitude of the  implementation of major government  public restrictions designed to mitigate the severity of the epidemic;
(2) the importance of both reported and unreported cases in interpreting the  number of reported  cases; and
(3) the importance of asymptomatic infectious cases in the disease transmission.
In our model formulation, we divide infectious individuals into asymptomatic and symptomatic infectious individuals. 
The symptomatic infectious phase is also divided into reported and unreported cases. 
Our model formulation  is based on our work \cite{LMSW}, in which
we developed a method to estimate epidemic parameters at an early stage of an epidemic, when the number of cumulative cases grows exponentially. The general method in \cite{LMSW}, was applied to the COVID-19 epidemic in Wuhan, China, to identify the constant transmission rate corresponding to the early exponential growth phase.

In this work, we use the  constant transmission rate in the early 
exponential growth phase of the COVID-19 epidemic identified in \cite{LMSW}. 
We model the effects of the major government imposed  public restrictions in China, beginning on January 23, 
as a time-dependent exponentially decaying transmission rate after January 24.
With this time dependent exponentially  decreasing transmission rate, we are able to fit with increasing accuracy, our model simulations  to the Chinese CDC reported case data for all of China, forward in time through February 15, 2020.

Our model demonstrates the effects of implementing major government public policy measures. By varying the date of the implementation of these measures in our model, we show that had implementation  occurred one week earlier, then a significant reduction in the total number of cases would have resulted. We show that if these measures had occurred one week later, then a significant increase in the total number  of cases would have occurred. We also show that if these measures had been less restrictive on public movement, then a significant increase in the total size of the epidemic would have occurred.  It is evident, that control of a COVID-19 epidemic  is very dependent on an early implementation and a high level of restrictions on public functions. 

We varied the fraction of unreported cases involved in the transmission dynamics. We showed that if this fraction is higher, then a significant increase in the number of total cases results. If it is lower, then a significant reduction occurs. It is evident, that control of a COVID-19 epidemic  is very dependent on identifying and isolating symptomatic unreported infectious cases. We also decreased the parameter $\nu$ (the reciprocal of the average period  of asymptomatic infectiousness), and showed that the total number of cases in smaller. It is also possible to decrease $\eta$ (the reciprocal of the average period of unreported symptomatic infectiousness), to obtain a similar result. It is evident that understanding of these periods of infectiousness is important in understanding the total number of epidemic cases. 

Our model was specified to the COVID-19 outbreak in China, but it is applicable to any outbreak location for a COVID-19 epidemic.

\end{document}